\documentclass[acmsmall,screen]{acmart}
\AtBeginDocument{%
  }

\setcopyright{acmlicensed}
\copyrightyear{2018}
\acmYear{2018}
\acmDOI{XXXXXXX.XXXXXXX}

\acmJournal{JACM}
\acmVolume{37}
\acmNumber{4}
\acmArticle{111}
\acmMonth{8}

\begin{document}

\title{Rethinking Higher Education:
From Fixed Curricula to Learnity Graphs}

\author{Smadar Szekely}
\authornote{}
\email{smadar.szekely@weizmann.ac.il}
\affiliation{%
  \institution{Weizmann Institute of Science}
  \city{}
  \country{Israel}
}

\author{Judith Gal-Ezer}
\authornotemark[1]
\email{galezer@openu.ac.il}
\affiliation{%
  \institution{The Open University of Israel}
  \city{}
  \country{Israel}
}

\author{David Harel}
\authornotemark[1]
\email{david.harel@weizmann.ac.il}
\affiliation{%
  \institution{Weizmann Institute of Science}
  \city{}
  \country{Israel}
  }

\renewcommand{\shortauthors}{Szekely et al.}

\begin{abstract}
Higher education stands at a turning point. In an era where knowledge is increasingly accessible and which is, more often than not, mediated by advanced Artificial Intelligence (AI), the value of traditional curricula models warrants reconsideration. This does not imply that one should replace thorough academic studies. Universities remain essential in providing foundational knowledge, theoretical depth and conceptual grounding. The challenge is to extend these educational facets with learning environments that foster creativity, interdisciplinary integration, hands-on experience, and especially long-term development. 	
In this paper, we introduce a lifelong learning framework that integrates academic, professional, and personal learning, centered on a new concept that we term learnity graphs, a structured representation of learning as interconnected units of knowledge, skills, experience, and actual artifacts, coupled with a method for presenting, and leveraging it.

\end{abstract}

\ccsdesc[500]{Applied computing~Education}

\keywords{learnity graph, higher education, generative AI, personalized learning, lifelong learning, learning pathways, knowledge graphs}

\maketitle

\section{Introduction}
Current models of higher education promote the provision of strong foundations, but remain largely structured around predefined programs and fixed curricular paths. This paper suggests that such structures are insufficient for representing the full scope of how individuals develop knowledge, skills, and experience across academic, professional, and personal contexts, particularly in the era of AI. This limitation motivates the need for a more flexible and integrative representation of learning.
\subsection{The Current Structure of Higher Education Degrees}
Higher education systems are commonly organized into three levels: undergraduate (bachelor’s), graduate (master’s), and doctoral programs. Undergraduate education provides basic knowledge, graduate programs offer specialization and advanced study, and doctoral programs focus on research and the creation of knowledge, and the skills these require.
This three-cycle model was formalized in the year 1999 in the Bologna Declaration, shaping higher education systems globally and supporting comparability and mobility across institutions \cite{bologna1999}.

In computing, these structures are well documented. Computing Curricula 2020 (ACM/IEEE) emphasizes the role of undergraduate programs in preparing students for professional practice and research \cite{cc2020}. Master’s programs serve diverse goals, including specialization and professional development \cite{cs2008}, while doctoral education develops the students’ ability to become part of the research workforce, which is essential for impactful innovation \cite{futurecs2022,rude2016}. While effective, this structure evolves slowly relative to rapidly changing knowledge domains and the associated methods and tools. When new fields or techniques emerge (e.g., data science and machine learning), institutions typically respond by creating new degree programs, often ones that overlap significantly with existing ones. Such responses preserve institutional structure but may not reflect deeper epistemic shifts. Emerging fields and methods often represent intersections of existing competencies rather than entirely new disciplines.

As noted in higher-education research \cite{barnett2004,wheelahan2015}, rigid degree structures struggle to keep pace with evolving knowledge, this often leads to program proliferation with only marginal differentiation.
\subsection{Toward Flexible Learning Architectures}
An alternative is to view emerging fields as evolving clusters within a broader learning network. Rather than creating new degrees, institutions could support the dynamic emergence of such clusters within a flexible learning architecture.

We thus propose a shift from viewing curricula as a collection of fixed units toward a dynamic architecture of skills, relationships, and experiences that evolves over time. The basic unit of this architecture is what we shall call the learnity – a learning entity representing knowledge, competencies, and/or experience. Learnities form the building blocks of learning pathways, and together with the relationships between them define a learnity graph –  a dynamic structure that enables flexible and evolving educational trajectories.

While micro-credentials decompose education into smaller certifiable units, our proposal focuses on representing learning as an evolving network of competencies, experiences, artifacts, and relationships.
\subsection{AI in Learning Development}
As access to information becomes abundant and to a large extent automated, the key challenge shifts toward applying knowledge, integrating ideas across domains, and developing meaningful experience. Educational value increasingly lies in developmental pathways rather than in information acquisition alone.

AI plays a central role in this shift, serving two complementary roles. First, rather than replacing learning, AI can act as a cognitive partner, supporting access to knowledge, enabling reflection, and assisting learners in structuring and deepening their understanding. Second, AI also enables recommendation mechanisms that provide personalized developmental guidance, based on a learner’s existing knowledge, professional experience, and a learning trajectory. Some research has highlighted the potential of AI to enhance personalization, metacognition, and adaptive learning \cite{holmes2019,luckin2016}. Within this context, the learnity graph will be seen to serve as a structural representation of evolving learning, enabling the identification of unique pathways and opportunities for growth.

This paper introduces the conceptual framework; building a full operational system remains a challenge for future interdisciplinary research.
\section{Related Work}
Prior work has examined curriculum design, competency-based education, lifelong learning, and the recognition of informal and workplace learning. More recently, growing attention has been given to generative AI, adaptive learning environments, and AI-supported personalized learning pathways \cite{denny2024,baylycastaneda2024,xu2025}. These directions emphasize learner agency, evolving competencies, and the need for educational models that support continuous and personalized development in rapidly changing knowledge environments. Collectively, these strands of research highlight the limitations of rigid program-centric educational structures and motivate the need for more flexible and dynamic representations of learning. Our work builds on these directions by proposing the learnity graph as a unified framework integrating knowledge, experience, artifacts, and development over time.

Academic systems traditionally emphasize formal credentials. However, professional experience often reflects aspects of cognitive depth, adaptability, and judgment that are not captured by degrees. Workplace learning research underscores the importance of experiential and informal learning \cite{eraut2004,billett2011}.  Similarly, research on learning trajectories (e.g.,\cite{pea1993}) emphasizes that learning unfolds across multiple contexts over time rather than within fixed institutional boundaries. The learnity-based framework proposed here integrates such experience into academic trajectories rather than treating it as peripheral.

Lifelong learning further challenges linear education and career models, emphasizing continuous development across the lifespan \cite{field2006}. Portfolio-based approaches document this growth through artifacts and reflection \cite{barrett2007}. In a learnity graph, such artifacts become structured components within a broader network. At the same time, competency-based approaches (e.g.,\cite{mulder2017}) decompose learning into smaller units, but often rely on predefined and relatively static structures.

Motivation research also supports flexible learning architectures. Self-Determination Theory highlights autonomy, competence, and relatedness as key drivers of engagement \cite{ryan2017,deci2000}. Learnity graphs can support these needs through personalized pathways, visible progress and collaborative structures.

While learning can be represented as interconnected networks without utilizing AI, recent advances in AI significantly enhance their usability. AI enables scalable search and navigation within large learning graphs, supports the identification of meaningful connections, and facilitates the forecasting of potential learning trajectories\cite{aburrasheed2024,presutti2025}. Recent work further emphasizes the role of AI-supported personalized learning environments that adapt to learners’ evolving needs and pathways \cite{denny2024,baylycastaneda2024,xu2025}.
Knowledge graph approaches (e.g., \cite{noy2001}) represent entities and their relationships, typically within structured domains, while critical perspectives (e.g., \cite{deburgh2012}) challenge fixed and linear representations, emphasizing more fluid and evolving configurations. A learnity graph builds on these directions by integrating the resulting insights into a unified model that captures the dynamic development of learning, combining knowledge, experience, and growth over time.
\section{Learnity Graphs}
We propose a framework that represents learning as a network of interconnected learning entities. A learnity is a minimal meaningful unit, which captures a capability, concept, experience, or demonstrated competence, emerging from coursework, projects or professional activity.

Learnities are connected through relationships, such as dependencies, interdisciplinary integration, composition and specialization. A set of learnities and their relationships form a learnity graph, where nodes represent learnities and edges represent relationships. This structure enables the modeling of learning as an evolving network rather than a fixed sequence.

For example, a programming learnity might be quite rich,  and could include knowledge (syntax, control structures), skills (implementation, debugging), artifacts (code, projects), and experience (collaboration, industry work). It may relate to other learnities in a learnity graph via edges (i.e., links) that represent prerequisites (e.g., discrete mathematics to algorithms), compositional links (programming + version control to software project), and interdisciplinary links (programming and data analysis). Across different learners, such graphs can be compared by structure and evidence, for example by the overlap of nodes (as shown in figure 4), types of relationships, depth of dependencies, and associated artifacts, enabling assessment of both similarity and distinctive developmental paths. Thus, a learner’s development is represented as a learnity graph, which may extend beyond traditional curricula and complement or replace conventional CVs.

Figure 1. shows an example of the  learnity graph of a second year student, Maya, combining academic and industry experience. Maya’s graph illustrates how her foundational academic learnities connect with her industry-based ones, highlighting the integration of knowledge, skills, artifacts, and experience into a unified developmental structure.

Figure 2 presents a  more general learnity graph, spanning academic foundations and a long-term professional trajectory. This structure illustrates how individual learning pathways emerge through the combination of knowledge, skills, artifacts, and experience across contexts. 

Figure 3 presents a learnity graph of a real estate lawyer with experience, and with professional development courses over and above his or her academic degree.

Each figure combines two complementary representations of the same learnity graph: a layered structural view organizing learnities by domain and developmental progression, and a relational graph view emphasizing connections between learnities. 

\begin{figure}
    \centering
    \includegraphics[width=0.8\linewidth]{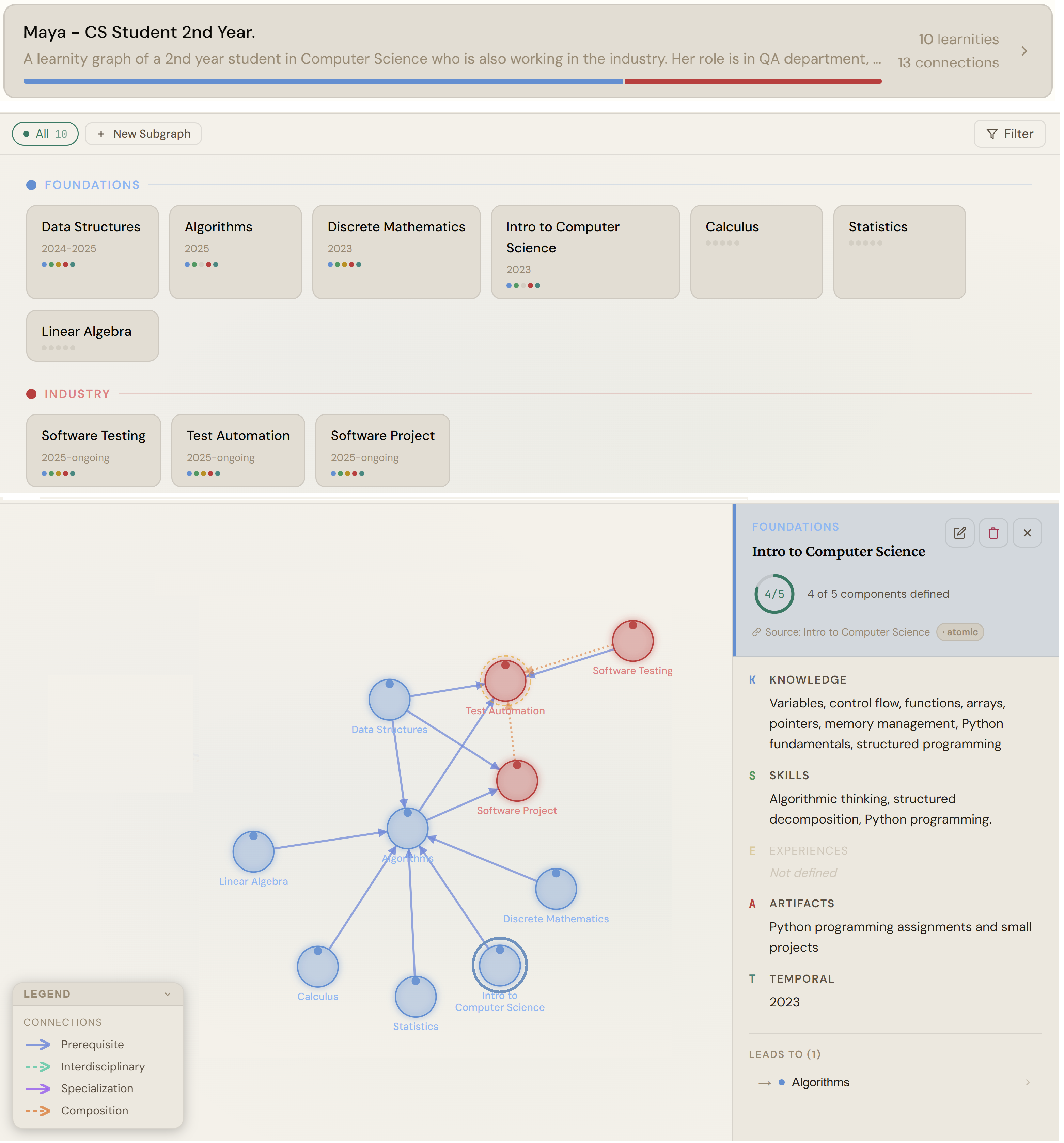}
    \caption{Example of a learnity graph for a second-year computer science student.}
    \label{fig:placeholder}
\end{figure}
\begin{figure}
    \centering
    \includegraphics[width=0.8\linewidth]{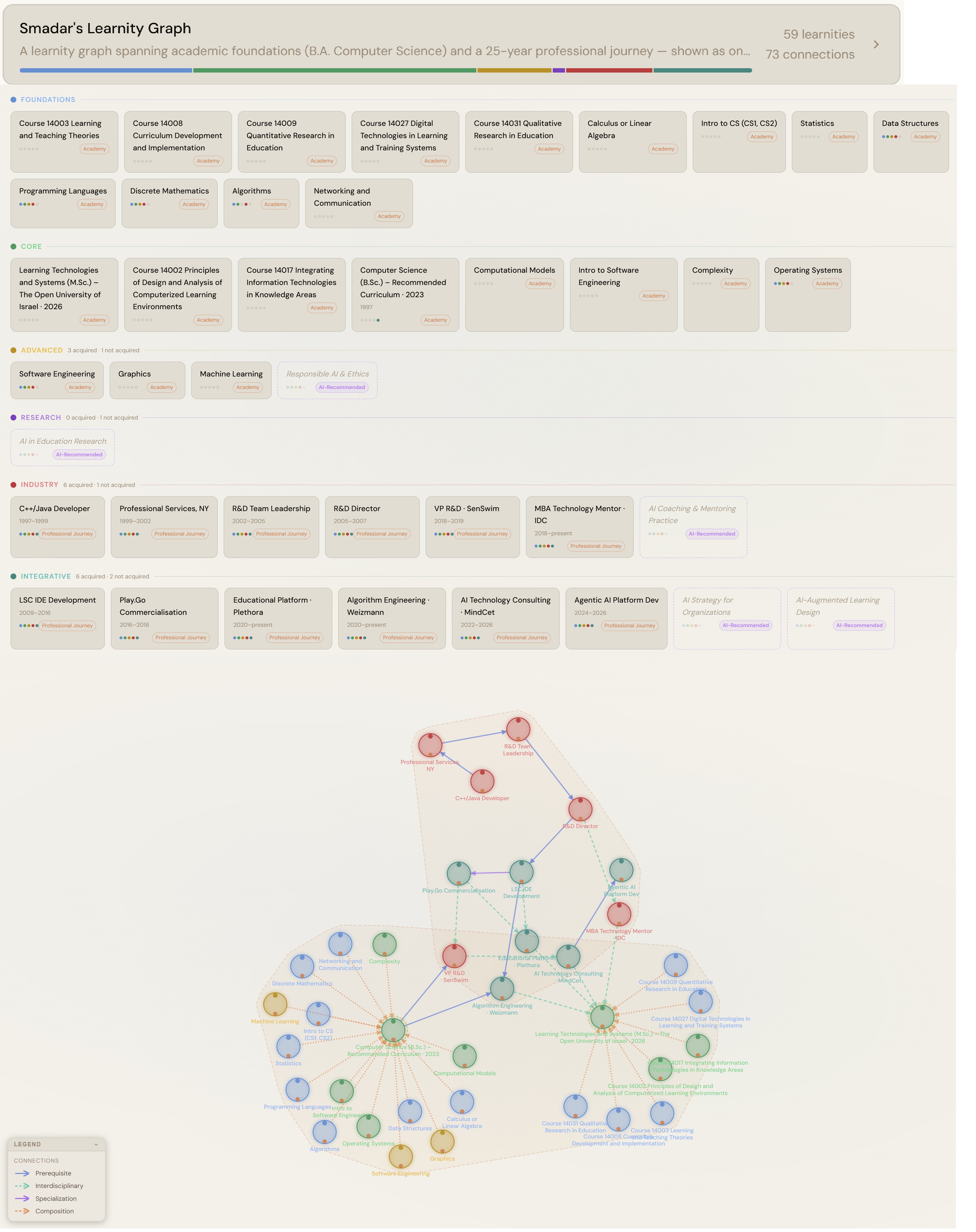}
    \caption{Example of a mature learnity graph. }
    \label{fig:placeholder}
\end{figure}
\begin{figure}
    \centering
    \includegraphics[width=0.8\linewidth]{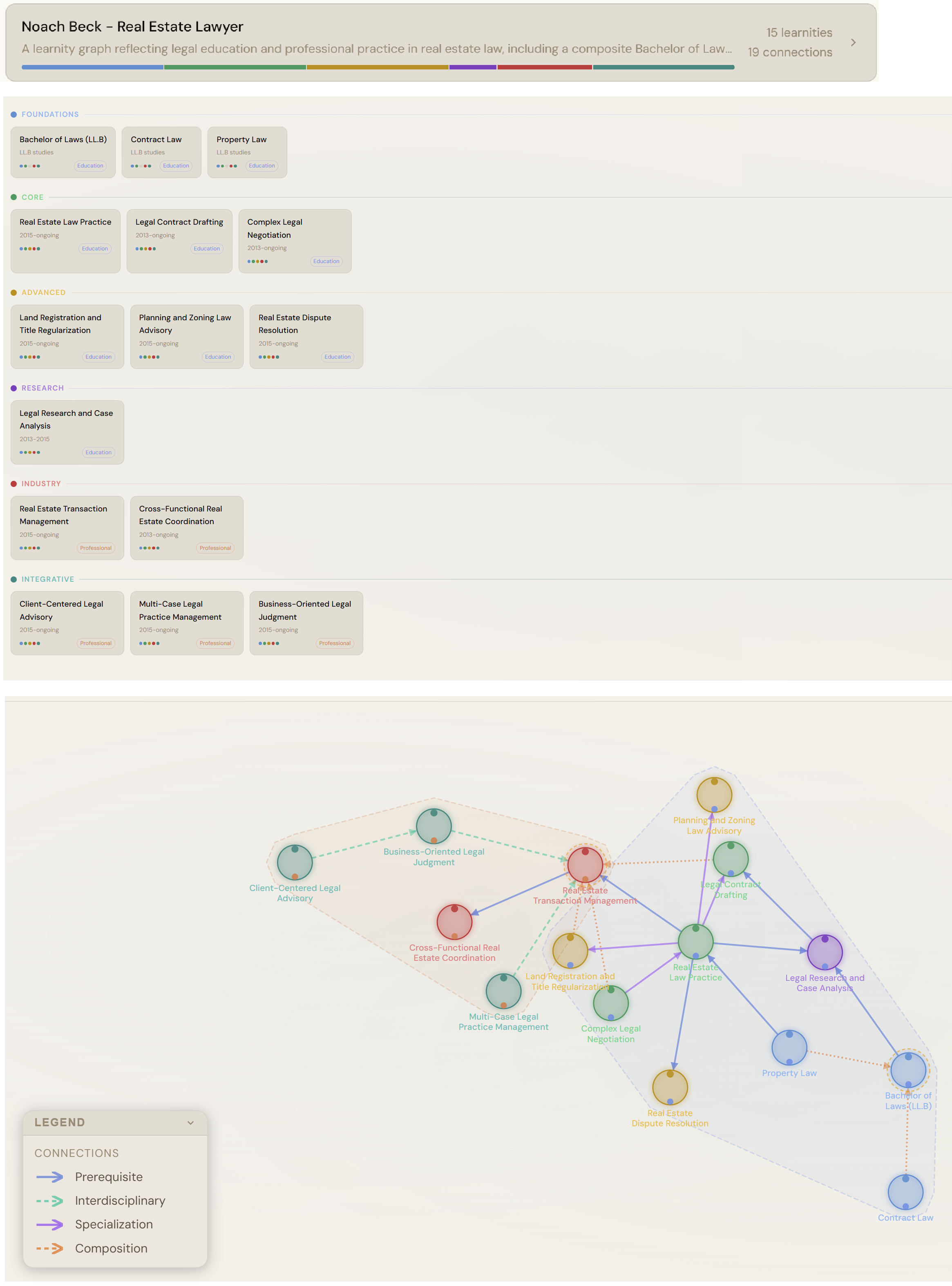}
    \caption{A learnity graph of a real estate lawyer, reflecting legal education and professional practice.
}
    \label{fig:placeholder}
\end{figure}
\begin{figure}
    \centering
    \includegraphics[width=1\linewidth]{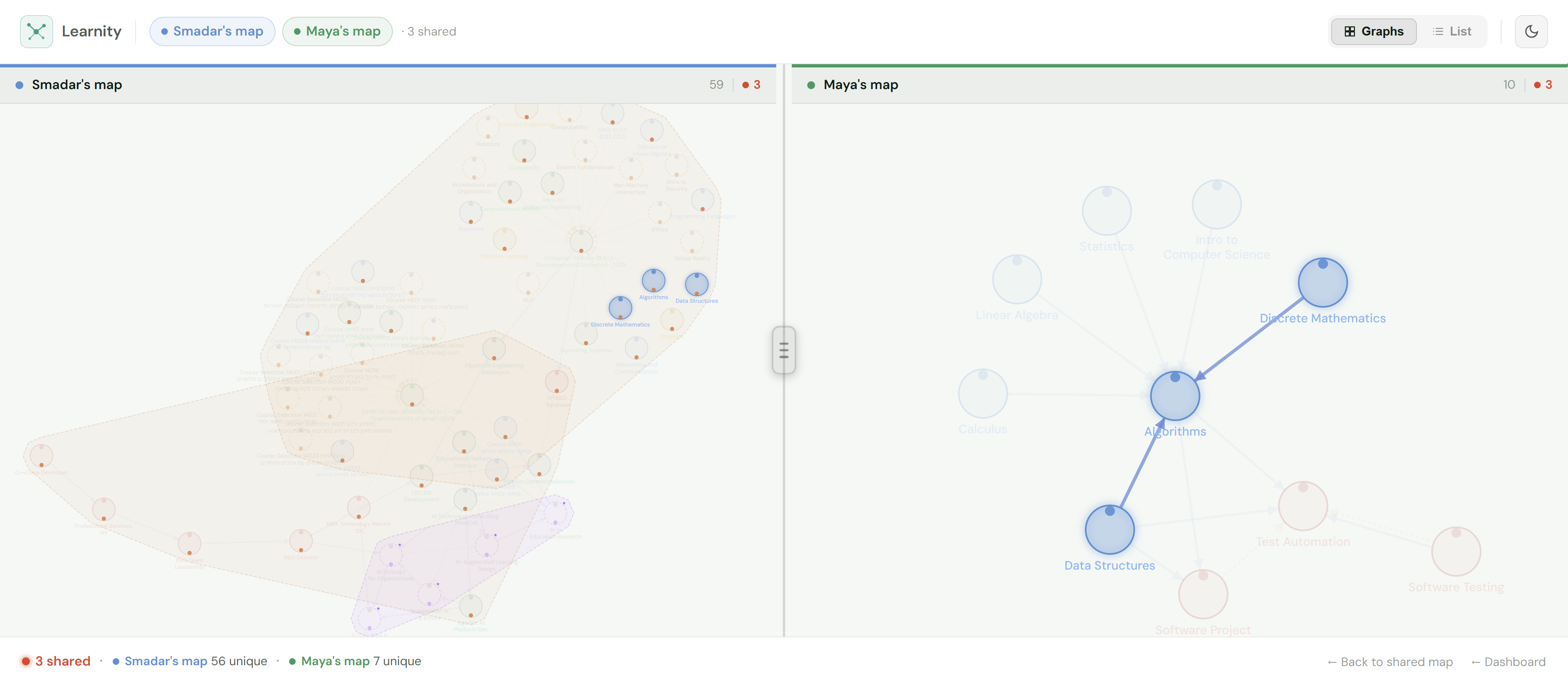}
    \caption{A comparison view that highlights the shared learnities of learners and demonstrates the potential of working with learnity graphs in assessing similarity and distinctive developmental pathways. 
}
    \label{fig:placeholder}
\end{figure}
\section{Design Considerations for Learnity Graphs}
\textbf{Granularity and Emergent Learnities.} While some learnities represent foundational capabilities, others may emerge from the integration of multiple learnities into higher-level competencies. Defining when a learnity should be treated as a distinct node in the learnity graph is critical to avoiding graph inflation and preserving interpretability. Clear criteria are therefore required for creating, aggregating, and validating learnities within the graph.

\textbf{Types of Relationships.} Relationships between learnities, depicted as edges in the graph, capture different forms of learning structure. Common types include prerequisite dependencies, interdisciplinary interactions, specialization pathways, and compositional integration. Distinguishing between these relation types enables richer modeling of learning processes and supports more precise analysis of development and transfer across domains.

\textbf{Structural Dimensions.} Layers and Subgraphs. Learnity graphs may be organized hierarchically admitting multiple structural dimensions. Layers can represent depth or level of development (e.g., foundational to advanced), while subgraphs can represent context (e.g., academic, professional, or self-directed learning). Such dimensions should be designed to be orthogonal and to provide complementary perspectives on a learner’s development.

\textbf{Clarity and readability.} Constructing learnity graphs may seem complex, but similar challenges arise in graph visualization. There is a large body of work on graph visualization, by which  the two-dimensional layout of complex graphs can be carried out in a way that enhances clarity and readability, suggesting that learnity graphs can be effectively structured and interpreted (see, e.g., \cite{tamassia2013,harel2002}).

\section{Implications}
The learnity graph introduces a shift in how learning and competence are represented. Each learner develops a unique graph reflecting individual experiences, artifacts, and development. In a world where knowledge is increasingly accessible and shared, value shifts from the possession of knowledge to the unique structure of an individual’s learning trajectory. Such uniqueness becomes a source of innovation, as novel combinations of knowledge, experiences, and perspectives create the conditions for creative problem solving and the emergence of new ideas. Recent discussions of AI-supported learning ecosystems further highlight the importance of adaptive, interdisciplinary, and continuously evolving learning configurations \cite{denny2024,baylycastaneda2024,xu2025}.

To support cross-institutional use, a shared representational language and standardization are required, including definitions of learnities and their relationships, development levels, contexts, and evidence types. However, standardization must remain flexible. Overly rigid structures risk reproducing the limitations of traditional curricula. Effective frameworks should guide rather than constrain \cite{mulder2017}. Figure 5 shows our proposed learning ecosystem.

\begin{figure}
    \centering
    \includegraphics[width=1\linewidth]{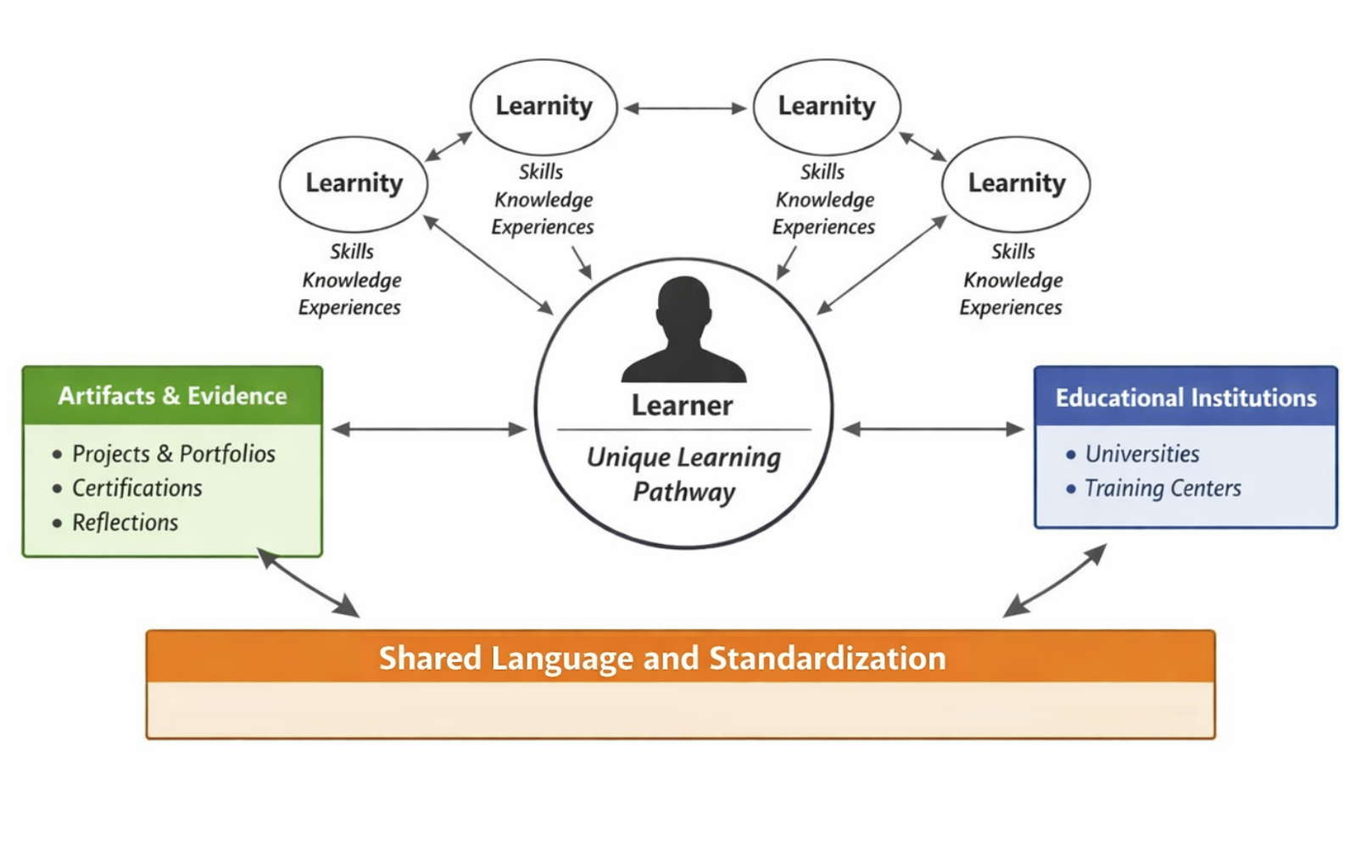}
    \caption{The  Learning Ecosystem
}
    \label{fig:placeholder}
\end{figure}
Implementing this ecosystem requires new infrastructure and raises significant challenges, including knowledge representation, evaluation mechanisms, technological scalability, and governance for cross-institutional recognition. While this paper introduces the concept, we leave full implementation for future work, which would clearly require interdisciplinary collaboration.

Nevertheless, initial prototype systems can already provide valuable insight into how such an ecosystem may operate in practice. We provide an example of an initial infrastructure demonstrating how the learnity-graph concept may be realized in practice (\href{https://learnity-graph.vercel.app}{https://learnity-graph.vercel.app}). This system enables exploration of both the opportunities and challenges of representing learning as evolving graphs, including personalization, development tracking, learning recommendation, and issues of structure, standardization, and interpretation. 

\section{Conclusion}

As generative AI systems increasingly support personalized guidance, recommendation, and knowledge navigation, frameworks capable of representing learning as dynamic and evolving networks may become increasingly important for future educational ecosystems \cite{aburrasheed2024,presutti2025}. The learnity graph offers one possible conceptual direction for integrating foundational academic study, lifelong learning, and AI-supported exploration within a unified representation of human development.

Several key questions remain:         	
How can learnity graphs be validated and standardized?          	
Who grants legitimacy to learnity portfolios?	
How can the learnity approach avoid replicating traditional grading systems?   How should algorithmic guidance be balanced with human choice?       	
More common challenges such as: privacy, data ownership, algorithmic transparency, and evidence validation should also be addressed.

A natural next step is to develop a suitable infrastructure to support learnity graphs, potentially through LMS systems centered on learnity graphs rather than on courses. Core components include a graph engine, an evidence layer, an analytics layer, and a recommendation engine. Figure 6 presents an example of such a system.

\begin{figure}[htbp!]
    \centering
    \includegraphics[width=0.8\linewidth]{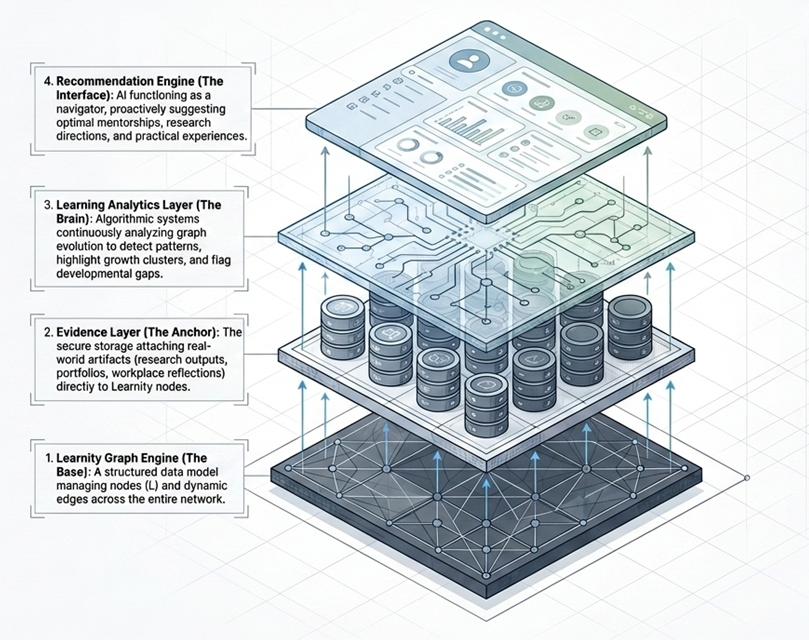}
    \caption{Learnity graph infrastructure. (Image generated by NoteBookLM.)
}
    \label{fig:placeholder}
\end{figure}
Future research is needed to explore the implementation of the approach into systems that can operate across institutions, enabling lifelong learning records that transcend institutional boundaries.

\clearpage
\bibliographystyle{ACM-Reference-Format}
\bibliography{sample-base}











\end{document}